# Prediction of the electron density of states for crystalline compounds with Atomistic Line Graph Neural Networks (ALIGNN)

Prathik R. Kaundinya[1], Kamal Choudhary[2,3], and Surya R. Kalidindi[1,4*]

[1]School of Computational Science and Engineering, Georgia Institute of Technology, Atlanta, Georgia 30332, USA.

[2]Materials Science and Engineering Division, National Institute of Standards and Technology, Gaithersburg, Maryland 20899, USA.

[3]Theiss Research, La Jolla, CA 92037, USA.

[4]G.W. Woodruff School of Mechanical Engineering, Georgia Institute of Technology, Atlanta, Georgia 30332, USA.

[*]Corresponding author: surya.kalidindi@me.gatech.edu

## Abstract

Machine learning (ML) based models have greatly enhanced the traditional materials discovery and design pipeline. Specifically, in recent years, surrogate ML models for material property prediction have demonstrated success in predicting discrete scalar-valued target properties to within reasonable accuracy of their DFT-computed values. However, accurate prediction of spectral targets such as the electron Density of States (DOS) poses a much more challenging problem due to the complexity of the target, and the limited amount of available training data. In this study, we present an extension of the recently developed Atomistic Line Graph Neural Network (ALIGNN) to accurately predict DOS of a large set of material unit cell structures, trained to the publicly available JARVIS-DFT dataset. Furthermore, we evaluate two methods of representation of the target quantity - a direct discretized spectrum, and a compressed low-dimensional representation obtained using an autoencoder. Through this work, we demonstrate the utility of graph-based featurization and modeling methods in the prediction of complex targets that depend on both chemistry and directional characteristics of material structures.

# 1. Introduction

Although physics-based modeling approaches such as density-functional theory (DFT) have been used extensively in the development of material structure-property relationships, they generally incur a high computational cost. As such, they are not well suited to materials discovery efforts that demand high-throughput screening to identify novel chemistries and material structures possessing certain desired combinations of physical and chemical properties. In this context, Machine Learning (ML)-based surrogate models trained on available DFT datasets offer an attractive avenue for the rapid screening of the extremely large materials design space. One of the primary potential applications of high-throughput screening with ML-based surrogate models is in the identification of materials that are well suited for use in energy storage applications. Materials selected for such applications generally have a desired stability (i.e., target formation energy range), exhibit a particular conductive/insulating capability (i.e., target electronic bandgap range) and have a particular specific heat capacity (i.e., phononic and electronic heat capacity ranges). Thus, a suitable workflow for identifying such materials is to shortlist a set of candidate materials (derived from the extensive materials search space) and perform DFT calculations for even more accurate property prediction on this shortlisted set of candidate materials.

ML-based surrogate models have been successfully employed in predicting scalar bulk material properties such as optical and electronic bandgaps [1], formation energy [2], Debye temperature [3], atomization energies [4] and polarizability of crystalline compounds [5]. ML models have also been developed to predict electronic properties such as the position of Wannier centers [6], electron charge density [7], and electron density of states [8]. These electronic properties can then be used to predict other physical properties of the material (e.g., the estimation of exchange-correlation energy [9] from the electron density). Of particular interest in this study is the electron Density of States (DOS), which reflects the number of states that may be occupied by the electrons in the material at a spectrum of energy levels. The DOS can be used to compute directly many other useful physical properties of the material, such as the effective mass of electrons in charge carriers [10], the electronic contribution to heat capacity in metals [11], and the electronic bandgap of crystal structures. The salient features of the DOS spectrum include the positions and magnitudes of prominent peaks, and the value of the DOS at the Fermi energy (i.e., at the origin of the plot).

Much of the prior work in ML-based predictions of the DOS spectra of materials has been limited to predicting certain features of it, learning patterns in the DOS across similar compounds or restricting analysis to datasets with limited chemical diversity. This is largely due to the complexity arising from the target being a spectrum (as opposed to a scalar-valued quantity) and the availability of limited training data. For example, Schutt et al. compared two different information encoding schemes (the Coulomb matrix and Partial Radial Distribution Function (PRDF)) and proposed a model using kernel ridge regression to predict the value of the DOS at the Fermi energy ($E_f$), which is closely related to the Seebeck coefficient and the critical temperature of superconductors [12]. As another example, Broderick and Rajan extracted the main features of the DOS spectra of single-element crystal systems with Principal Component Analysis (PCA) and identified sixteen spectral patterns that could be weighted differently to yield close approximations to the DOS spectra of transition metals [13]. Another effort demonstrated a method for "pattern learning" the DOS spectra of multi-component alloy systems [8b] and nanoparticles [8c]. In these studies, PCA was used to generate weighted components of several basis spectra, which were then used to adequately represent the DOS spectra of the different material systems. In a more sophisticated approach, Chandrasekaran et al. [7c] developed a rotationally invariant grid-based fingerprinting scheme using Gaussian functions of varying widths at each energy level. This representation was used to model the Local Density of States (LDOS) at each energy level within polyethylene and aluminum crystal systems, which were then summed to produce the overall DOS spectra. Mahmoud et al. [8a] compared different approaches for representing the DOS (with principal components, a pointwise discretization or cumulative distribution function (CDF)) and developed predictors for the LDOS of several configurations of silicon.

The prior studies discussed above are not readily extendable to ML-based predictions of DOS for large collections of diverse crystalline materials exhibiting different symmetries and chemical elements. This difficulty stems from the need to develop an efficient featurization scheme for the diverse material structures (i.e., inputs) present in such large collections. It is important to note that electronic properties such as the DOS are sensitive to spatial and directional characteristics of the material structure (such as interatomic distances, bond angles and localized distortions). As such, the first step in a broadly applicable ML-based framework to predict the DOS involves developing a compact (but comprehensive) representation that accounts for the diverse structural and chemical information needed to represent complex material structures.

Typically, such representations should seek to quantify the salient details of the local atomic neighborhoods as the main features (i.e., regressors) controlling the complex interatomic interactions underlying the computations of the DOS. Some of these representations are kernel-based, accounting for individual interactions between pairs of atoms or a collection of atoms belonging to selected species (e.g., Smooth Overlap of Atomic Positions (SOAP) power spectrum [14], PRDF [12], and grid-based descriptors proposed by Chandrasekaran et al. [7c]). In an alternate approach, the molecular structure is comprehensively feature engineered using the framework of n-point spatial correlations [15], which utilize voxelized representations and benefit from the computational efficiency of the Fast Fourier Transform (FFT) algorithm [2e]. This framework aims to capture comprehensively the salient features in the molecular structure, including the directional information (e.g., bond angles) that is lost in the pairwise descriptors mentioned earlier.

Recently, a new approach using graph embeddings has been proposed to capture the short-range and long-range atomic interactions in molecular structures. Specifically, the use of crystal graphs (CG) as inputs to a Graph Convolutional Neural Network (GCNN) has been termed as CGCNN (Crystal Graph Convolutional Neural Network) [16]. Multiple variants of CGCNN have already shown promise for materials discovery problems (e.g., iCGCNN (improved CGCNN) [17], OGCNN (Orbital GCNN) [18] and GATGNN (Graph Attention GNN) [19]). In these approaches, the structure is typically represented as a graph, with the nodes capturing information about the individual constituent atoms and the edges encoding interatomic information (e.g., interatomic distances). The node representations are transformed over several layers (referred to as *graph convolution layers*) utilizing information from its close neighbors in the graph. Finally, the transformed node representations are aggregated and further transformed in a feed-forward neural network comprising of fully-connected (FC) layers, with the target property of interest being the output of these layers. These models have been successfully applied to predict the scalar properties of molecular structures with good accuracy. However, these early efforts did not capture directional information in crystal systems, and instead utilized limited rotationally invariant features such as interatomic distances, bond valences, and atomic radii [16-18]. The Atomistic Line Graph Neural Network (ALIGNN) [20] is a recently developed extension of GCNN approach that addresses the shortcomings described above by capturing interatomic distances and bond angles in two separate graphs known as the crystal graph and line graph, respectively. As a result of being able to capture the directional aspect of the local environment, ALIGNN models perform significantly better than CGCNN in the prediction of properties mentioned above (e.g., 50% improvement in the prediction of the formation energy of crystals, 30% improvement in the prediction of the electronic bandgap [20]).

In this study, we utilize the open-source implementation of the ALIGNN framework to effectively capture unit cell structural information of a broad range of crystalline materials (comprising of different structural symmetries and chemical elements) and predict their corresponding DOS spectra. Since the regression target is a spectral quantity (i.e., DOS), we also evaluate two methods of representation of the output: (i) a primitive discretization of the DFT-computed spectrum with a vector of 300 evenly spaced points in the energy range of ($-5$ to $10$ eV), and (ii) an autoencoder network to "learn" a concise low-dimensional representation of the high-dimensional target. In the second representation, the low-dimensional output of ALIGNN is passed through the decoder segment of the autoencoder network to recover the desired target (i.e., DOS). High-fidelity data for $\approx 56k$ crystalline materials was obtained from the publicly available Joint Automated Repository for Various Integrated Simulations – Density Functional Theory (JARVIS-DFT) dataset [21] and used to train the models described in this work.

## 2. ALIGNN Modeling framework

As mentioned earlier, ALIGNN encodes crystal structure information in the form of a crystal graph and a separate line graph. The crystal graph is constructed in the same manner as the CGCNN [16], and is comprised of two parts: i) a set of $I$ nodes representing the individual atoms present in the crystal unit cell, and ii) edges that quantify the connectivity between the nodes using suitably defined measures. Each node is associated with a vector embedding $\boldsymbol{v}_i$ ($i \in I$), which is formed by concatenating the following five attributes of the specific atom: electronegativity, electron affinity, number of valence electrons, first ionization energy, and atomic radius. Each edge, between a pair of atoms indexed by $i$ and $j$, is associated with another vector embedding $\boldsymbol{u}_{ij}$. In this work, the edges embed a simple scalar Euclidean distance measure between the atomic centers. Only the first sixteen nearest neighbors of each atom included in the crystal graph are considered in this work.

The line graph $\mathcal{L}$ is constructed from the crystal graph $G$ described above. Each node in the line graph corresponds to an edge in the crystal graph. In other words, the node embeddings in the line graph and the edge embeddings in the crystal graph share the same latent representation. Edges are then drawn between these nodes when a common atom is shared (e.g., between $u_{ij}$ and $u_{jk}$), with the edge embedding denoted by $t_{ijk}$ and reflecting the bond angle cosine for the interatomic angle formed between the ordered triplet of atoms indexed by $\langle i, j, k \rangle$. In other words, for each triplet of nodes (represented by the node indices $v_i$, $v_j$ and $v_k$, and edge vectors $u_{ij}$ and $u_{jk}$ in the crystal graph), the corresponding edge vector in the line graph, $t_{ijk}$, captures an angular measure of the bond angle involved.

As mentioned earlier, the node and edge representations are updated sequentially over several *graph convolution* layers. Specifically, the update consists of iterative modifications of the line graph and crystal graph, in order. Each update is known as an *edge-gated graph convolution*, and operates on a node and its local environment. The update process is described next for the line graph, with a similar procedure applied to update the crystal graph. Let $l$ index the updates of the line graph, operating on nodes $u_{ij}$ and their edges $t_{ijk}$. We start by computing normalized edge contributions as

$$\hat{t}_{ijk}^l = \frac{\sigma(t_{ijk}^l)}{\sum_{m \in K} \sigma(t_{ijm}^l) + \varepsilon} \tag{1}$$

where $\sigma(\cdot)$ indicates the sigmoid function, and $\varepsilon$ denotes a small constant added for numerical stability (taken as $1e-6$). The node vector of the line graph is then updated as

$$u'^l_{ij} = \Lambda_1^l u_{ij}^l + f\left(\sum_{j \in J} \hat{t}_{ijk}^l \odot \Lambda_2^l u_{ij}^l\right) \tag{2}$$

in which $\Lambda_1$ and $\Lambda_2$ are trainable weight matrices, $f$ indicates a nonlinear activation (such as the ReLU or SiLU) and $\odot$ indicates the elementwise multiplication operation. Note that the same weight matrix $\Lambda_2$ is shared between all neighbors in the second term of Eq. (2). $u'^l_{ij}$ represents an intermediate update of the node vectors of the line graph using only the information from its neighbors in it. The update from $u'^l_{ij}$ to $u_{ij}^{l+1}$ occurs with graph convolution over the crystal graph, as will be described later. Each edge in the line graph $t_{ijk}$ is updated as

$$t_{ijk}^{l+1} = t_{ijk}^l + f(\Omega_1^l u'^l_{ij} + \Omega_2^l u'^l_{jk} + \Omega_t^l t_{ijk}^l) \tag{3}$$

where $\boldsymbol{\Omega_1}$, $\boldsymbol{\Omega_2}$ and $\boldsymbol{\Omega_t}$ represent trainable weight matrices. As mentioned earlier, the update steps for the crystal graph are similar to those used for the line graph. For the crystal graph, the normalized edge contribution as applied in Eq. (1) yields the normalized intermediate edge vector $\widehat{\boldsymbol{u}}_{ij}^{\prime l}$. The normalized intermediate edge vector is then used to update the node vector of the crystal graph as $\boldsymbol{v}_i^l \rightarrow \boldsymbol{v}_i^{l+1}$, in a manner similar to Eqs. (2) and (3). The overall ALIGNN training schedule consists of $L$ graph convolution updates that are performed sequentially on the line graph and crystal graph, respectively. The final output of the crystal graph is obtained by performing an average pooling operation on all node vectors in it. The averaged node vector is then used as an input to a series of fully-connected (FC) layers to produce a prediction of the target. The architecture of the ALIGNN model used in this work is similar to the original model proposed by Choudhary and DeCost [20], extended to support predictions for a vectorial target (the DOS spectrum). Specifically, the implementation in this work entails two main extensions: i) an increase in the number of neurons in the output of the FC layers equal to the number of bins chosen for the vector target, and ii) the consideration of a larger number of neighbors for each atom included in the crystal graph (sixteen in this study vs twelve in the earlier study).

## 3. Application

As mentioned earlier, a dataset comprising $\approx$ 56k k crystal structures and their corresponding DOS spectra was obtained from the publicly available JARVIS-DFT repository and used to train an ALIGNN model. This dataset included crystal structural information (atomic centers, species, lattice constant, lattice type, etc.) and several DFT-computed chemical properties. Specifically, DOS spectra were generated for all crystal structures with the OptB88vdW functional [21c], with an automatic convergence for k-points and a Gaussian smearing of 0.01 eV. This dataset included crystalline compounds covering 89 species with varied chemistries. Figure 1 summarizes the frequency of occurrence of the different species in the compounds included in the dataset. It can be seen that there is a broad distribution of chemical species in the compounds included, with most species occurring in $\approx$ 3,000 compounds. Oxygen was the most frequently observed species (14,970 compounds) due to it being a part of several compound classes such as oxides, sulfates, carbonates and nitrates.

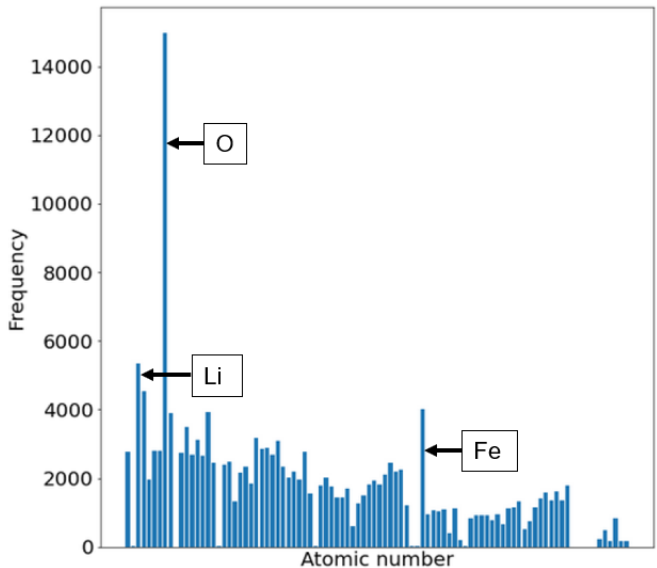

Figure 1. The frequency of occurrence of the different chemical species in the dataset used in this study.

The dataset was randomly partitioned into a $70\% - 10\% - 20\%$ split for use as train, validation (during training), and (fully blind) test sets, respectively. The primary purpose of the validation set was to implement an automated early stopping criterion during the model training phase. The input to the ALIGNN model comprised the chemical structure features (species and coordinates) and the target was the DFT-computed DOS. Figure 2 depicts the pipeline implemented to train the ALIGNN model. As shown in Figure $2(a)$, a nearest-neighbor search was performed for each atom to build the initial crystal and line graphs (similar to the original CGCNN [16]). Each atom was connected to its sixteen nearest neighbors, and the crystal and line graphs were initialized using the chemical and structural features described earlier in Section 2 (Figure $2(b)$). Following this, four ALIGNN edge-gated graph convolution updates were performed on both graphs, using the methodology detailed in Section 2 (Figure $2(c)$). Average pooling was performed on all nodes to yield a globally averaged node vector (Figure $2(d)$), which was further connected to four FC layers (Figures $2(e)$). Finally, the DOS spectrum was predicted as the output of the FC layers (Figure $2(f)$) using two different representations described in the next section.

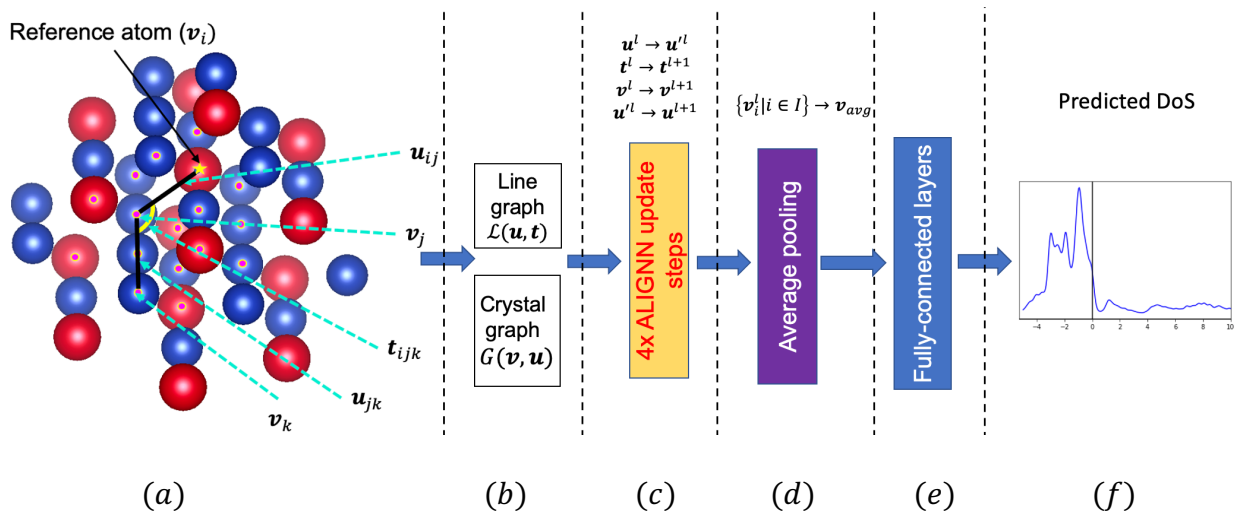

Figure 2. A flow diagram depicting the ALIGNN pipeline built in this study to predict the DOS spectrum of the compound. $(a)$ The atoms indicated by the pink marker represent the local neighborhood (i.e., connected atoms in the graph) corresponding to the reference atom. $(b)$ The embedding procedure creates the node and edge vectors of the crystal graph and line graph. $(c)$ The ALIGNN updates described in Section 2 are used to update the graphs. $(d)$ An average pooling operation is performed on the set of all node vectors present in the crystal graph $I$ to produce an averaged feature set. $(e)$ A series of fully connected neural networks are used to map the features to the target DOS spectrum. $(f)$ The predicted DOS spectrum is the output of the model.

## 4. Representation of the DOS

As mentioned earlier, since DOS is inherently a high-dimensional entity, it has presented significant challenges to prior machine learning efforts in literature. In this work, we explored two different representations of the DOS spectrum: i) a primitive discretization of the DOS using 300 uniform bins on the energy values, and ii) a low-dimensional representation of the DOS learned using an autoencoder network. Further details of these two approaches are presented next.

i. <u>Discretized representation of the DOS:</u> Since the DFT-computed DOS spectra were obtained with varying energy grids for each material, an interpolation scheme was necessary to standardize the representations for training and testing the ALIGNN models developed in this work. The discretized representation of the DOS was established by interpolating the DFT-computed DOS values between $-5$ and $10$ eV on a uniform grid with a spacing of $0.05$ eV.

This discretization interval was selected through multiple trials with the goal of achieving adequate resolution to unambiguously capture the salient features of the DOS spectra (e.g., major peak locations and their intensities) with the minimum number of bins. Specifically, each trial consisted of implementing the interpolation scheme with a different energy spacing on a set of randomly chosen candidate materials, and evaluating the reconstruction performance.

ii. Further, the DOS intensities were normalized by the total number of valence electrons in the crystal, thereby bounding the target DOS values to lie in the range [0,1]. The ALIGNN model was trained on the normalized DOS spectra.

iii. <u>Low-dimensional representation of the DOS:</u> As an alternate to the primitive discretized representation described above, we have also explored the utility of an autoencoder network for establishing high-value low-dimensional latent representations of the normalized DOS. This is because the 300 discretized values of the DOS described above are expected to exhibit some degree of correlation (i.e., dependency) amongst themselves. This observation is further supported by recent studies that have identified several patterns in the DOS spectra of different alloy compositions [8b, 8c]. Autoencoder networks have been shown to be ideal for addressing this task in other similar problems [22]. These networks typically consist of two connected components: (i) an encoder to map the high-dimensional input to a low-dimensional latent embedding, (ii) followed by a decoder that reconstructs the high-dimensional input data from the learned low-dimensional latent embedding. Figure 3 shows the architecture of the autoencoder network trained in this study. The encoder comprises of several fully connected layers, with a decreasing number of output neurons with each layer, as shown in the figure. Each layer of the encoder uses a ReLU activation function. The architecture of the decoder is designed to reverse the mapping of the encoder, with all layers comprising ReLU activation except for the last layer, which employs a sigmoid activation function to ensure that the reconstructed DOS spectrum only has values in the range of [0,1]. The autoencoder was trained on the same training data used to train ALIGNN, with the training objective set to minimizing the MSE (Mean Squared Error) between the decoder-reconstructed and the actual (input) DOS spectra.

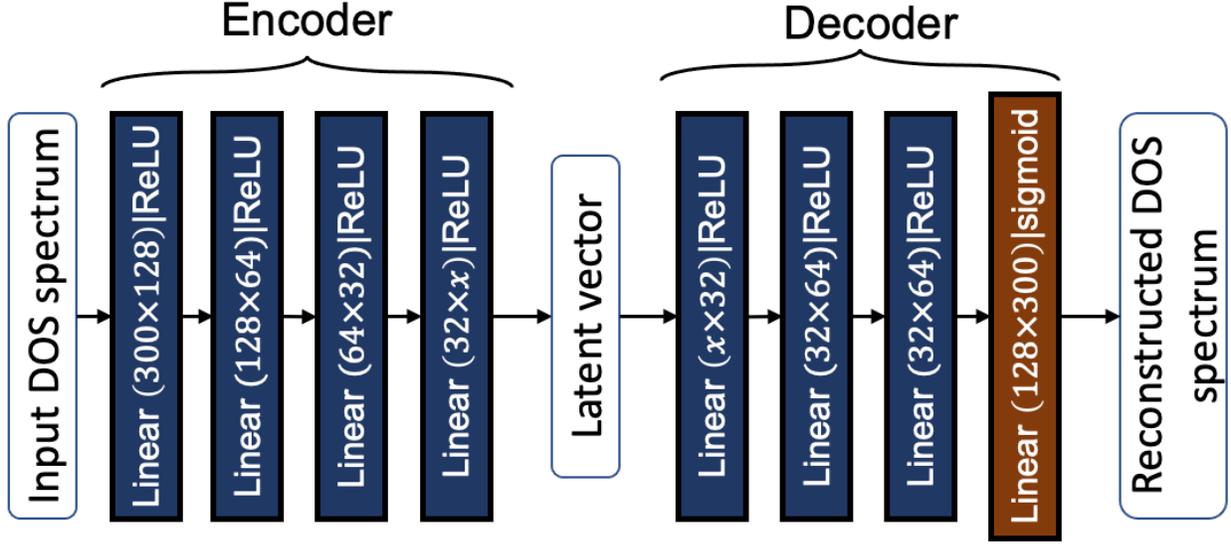

Figure 3. The architecture of the autoencoder used in this study. The encoder was designed to produce a low-dimensional representation of the 300-dimensional DOS vector, while the decoder was designed to fully reconstruct the 300-dimensional DOS vector from the low-dimensional representation.

5. **Results and discussion**

Separate ALIGNN models (referred to as Discretized ALIGNN (D-ALIGNN)) and autoencoder ALIGNN (AE-ALIGNN)) were trained for both representations of DOS presented in the previous section. The performance of the models on each test sample was quantified using two metrics – the mean absolute error ($MAE$) and the relative error ($RE$). These error metrics were computed using the predicted values of the DOS discretized over $N = 300$ points denoted as $\boldsymbol{p} \equiv \{p_1, p_1, \dots p_N\}$ and the corresponding DFT-computed values denoted as $\boldsymbol{a} \equiv \{a_1, a_1, \dots a_N\}$:

$$MAE = \frac{1}{N} \sum_{i=1}^{N} |p_i - a_i| \qquad (5)$$

$$RE = \frac{|\bar{p} - \bar{a}|}{\bar{a}} \qquad (7)$$

where $\bar{p}$ and $\bar{a}$ denote the mean DOS of the predicted and DFT-computed spectra. Based on the above metrics, a baseline prediction was made to compare with the performance of the trained models. The baseline prediction is defined as the mean DOS spectrum generated from all materials present in the training set. In other words, the baseline prediction corresponds to the prediction made by a model without any parameters or learning capability, simply as the mean of the training data. The metrics described above were also computed for the baseline prediction.

Figure 4 summarizes the predictive accuracy of the D-ALIGNN model for the test set. Figure 4($a$) presents a histogram of the MAE values for the test samples, with the shaded regions indicating different quartiles in the data. It can be observed that the DOS spectra for 92% of the test samples was predicted to within an MAE of 0.02 states/eV/electron, indicating the high accuracy of D-ALIGNN model. Most interestingly, the first two quartiles in Figure 4($a$) (comprising of 2,782 samples) exhibit predictive errors below 0.008 states/eV/electron and display reasonable agreement with the DFT-computed values. In addition, the average MAE over the entire test set was found to be 0.009 states/eV/electron, with only 6% of the materials having a predictive error greater than 0.02 states/eV/electron. The average MAE corresponded to a ~3.5 times improvement over the baseline model (which had an MAE of 0.031 states/eV/electron). Additionally, we observed a strong correlation of the higher prediction errors with the lack of adequate number of training points involving certain elements. Specifically, it was observed that the highest prediction errors occurred in compounds containing one of the following five elements: Cs, La, Ar, Ce and W. These specific elements were in less than 1.5% of the compounds included in the training set. However, the predictions are expected to improve as more data gets added to the training set. It was also noted that the average MAE values for the different classes of crystal structures (i.e., orthorhombic, cubic, hexagonal, monoclinic, triclinic, trigonal and tetragonal) were in a close range 0.007 to 0.009 states/eV/electron, indicating that the model produced in this work exhibits good predictions across all the crystal classes considered. This observation confirms that the implicit feature engineering in the D-ALIGNN model is capable of identifying the salient features across a diverse set of crystal structures. The histogram of the relative errors shown in Figure 4($b$) reaffirms the excellent predictive capability of the model, with 85% of the predictions exhibiting an RE under 0.2. In order to better visualize the predictions of the model, Figure 4($c$) depicts the DFT-computed and model-predicted DOS spectra for four random samples, one from each of the quartiles in Figure 4($a$). As seen from these comparisons, the general characteristics in the DOS spectra (such as peak locations and trends in the curve) are well captured by the model predictions. However, the model appears to sometimes predict non-existent peaks (as in $InH_2$) or understate the magnitudes of existing peaks (as in LaCo$_2$Ge$_2$). Additionally, the DOS at the Fermi energy (i.e., the y-intercept of the plot) appears to be computed accurately, with 90% of the materials having an absolute prediction error under 0.02 states/eV/atom. This allows for the accurate characterization of material's conductive nature (i.e., as insulators, semiconductors or conductors).

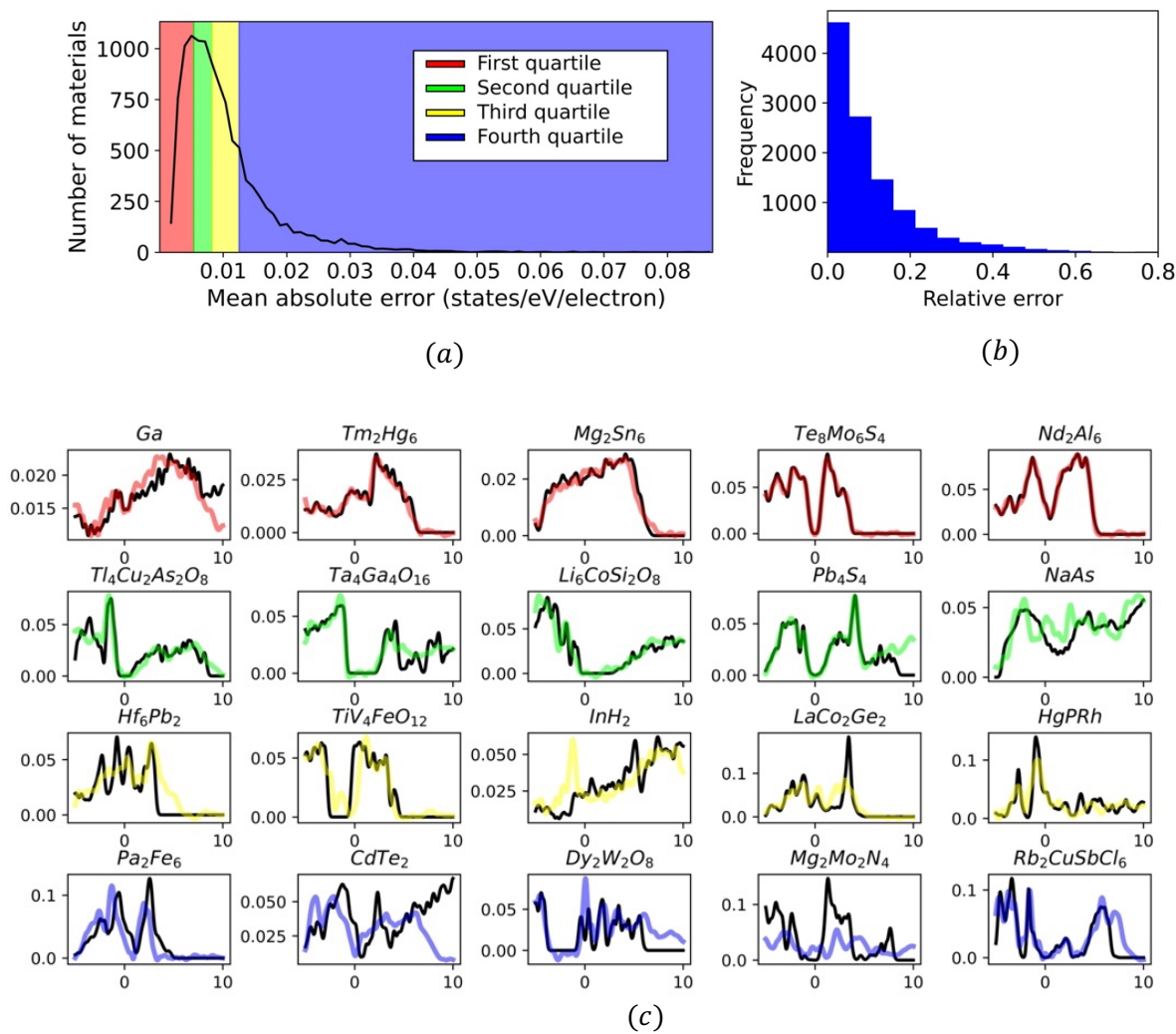

Figure 4. (*a*) The histogram of the MAE values for the D-ALIGNN predictions for the test set. The shaded colors represent four quartiles of the prediction error. (*b*) A histogram of the relative errors in the test set. (*c*) Four comparisons of the predicted DOS for randomly selected test points

from each quartile in $(a)$. In these comparisons, the actual DOS is shown as a black curve, and the predicted DOS is depicted by the colored curve matching the quartile color in $(a)$.

Table 1 shows the variation in the average MAE for the AE-ALIGNN model with the size of the encoded representation. It is important to note that the model complexity (i.e., number of parameters) is directly related to the size of the low-dimensional representation used as a target for the model (recall that the output is an FC layer with the same size as the target vector). As such, it is desirable to choose a latent dimension that is as small as possible to avoid a potential model overfit, but large enough to capture the complexity of the spectral output. As seen from Table 1, a latent vector of length 12 offers an optimal choice for adequately representing the DOS spectra in the dataset, since further improvement in the accuracy of the DOS prediction beyond this number of dimensions is fairly limited. Additionally, the average MAE of the AE-ALIGNN models appears to be higher than the D-ALIGNN model. This can be attributed to the loss in information through the compression achieved by the autoencoder. On average, the reconstruction MAE of the autoencoder compression for the case with a latent dimension of 12 was found to be 0.003 states/eV/electron, which contributed to the higher error in the AE-ALIGNN models than the D-ALIGNN model.

Table 1. The variation of the average MAE for the test predictions from the AE-ALIGNN model as a function of number of autoencoder features used for the representation of the DOS.

| Number of dimensions | Average MAE |
| --- | --- |
| 20 | 0.0112 |
| 16 | 0.0130 |
| 12 | 0.0134 |
| 8 | 0.019 |

Figure 5 depicts the results of the predictions of the AE-ALIGNN model with a target dimension of 12 on the materials in the test set. As seen from Figure 5$(a)$, the first quartile spans a larger range of MAE values in comparison with the D-ALIGNN model, indicating that the D-ALIGNN is more accurate for a larger number of samples. This reaffirms the hypothesis that the reconstruction error leads to loss of information and consequently a higher average MAE in prediction. Additionally, the relative error histogram depicted in Figure 5(b) is similar to the trend observed for the D-ALIGNN model, with 82% of the predictions of the DOS spectra exhibiting an RE under 0.2. The plots of DFT-computed and predicted DOS spectra of four randomly chosen samples, one from each of the quartiles in Figure 5(a), indicates that the AE-ALIGNN model does indeed provide good predictions. However, it is also clear that there is a slight loss in predictive accuracy of the AE-ALIGNN model compared to the D-ALIGNN model.

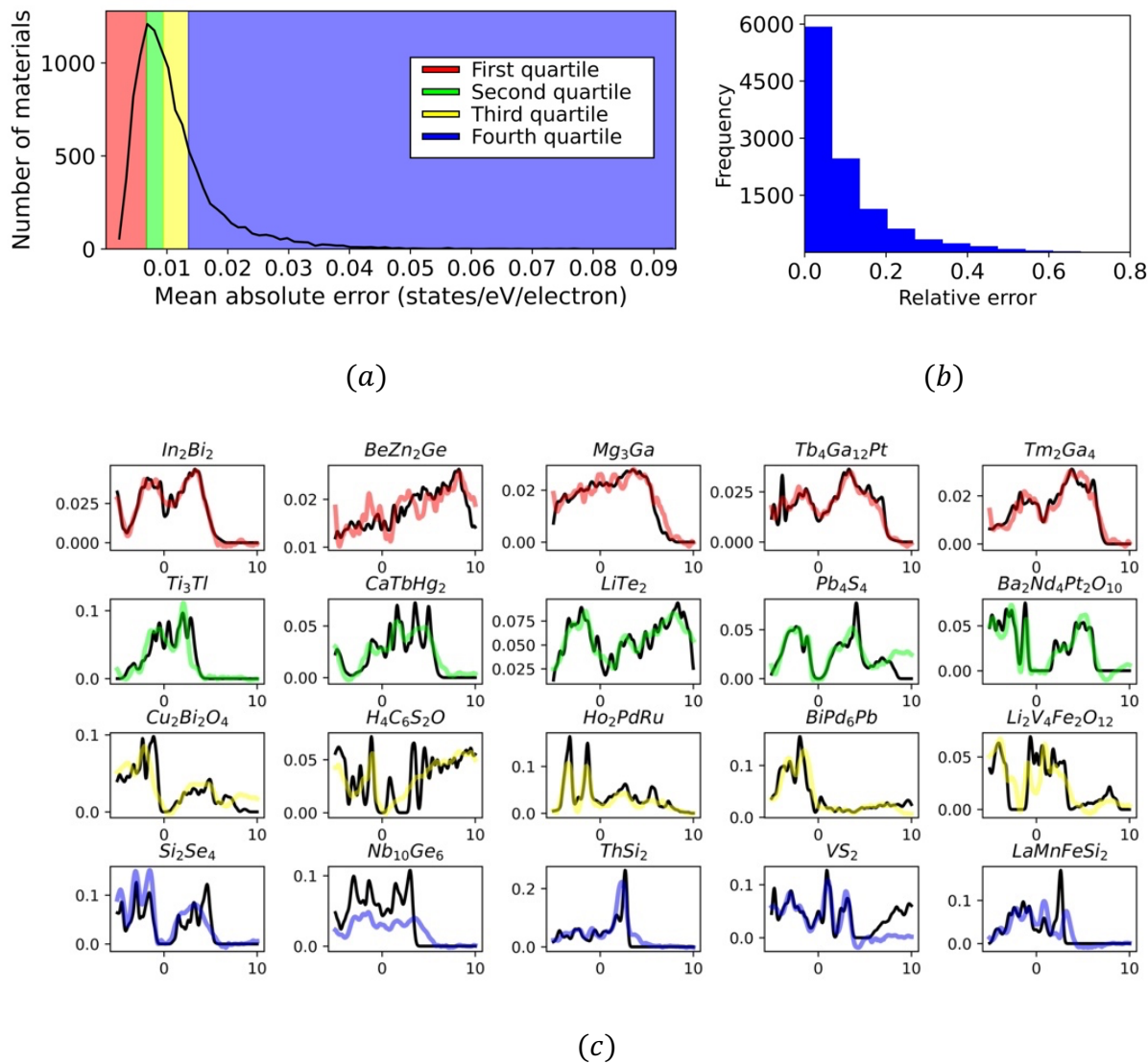

Figure 5. (a) The histogram of MAE values for the AE-ALIGNN predictions (with 12 latent dimensions) for the test set. (b) The histogram of relative errors in the test predictions. (c) Four comparisons of the AE-ALIGNN predicted and DFT-computed DOS spectra from each quartile shown in (a).

The models trained in this work can also provide useful insights into the electronic properties (that are derived from the DOS) of crystal structures and their relationships to the constituent species. For example, one of the primary criteria in selecting materials for photovoltaic applications is a desirable band gap range (usually between 1.4 to 1.6 eV). The band gap for a non-conducting material may be easily computed by calculating the energy difference between the first non-zero DOS values on either side of the y-axis in the DOS spectrum. In order to demonstrate the utility of the developed ALIGNN model to guide the search for materials with a required bandgap, we consider all binary compounds occurring in the test set and perform computations of the estimated bandgap using the predicted DOS spectra of the materials. Figure 6 depicts a heatmap of the periodic table, with each element displaying the average (non-zero) bandgap of all binary compounds containing the selected element. Note that elements not occurring in any binary compound in our dataset are shaded gray. In general, two major trends can be observed in the table: i) The average bandgap of binary compounds having one element from groups (i.e., columns) 1-2 and 16-18 appears to be relatively higher than those containing only transition metals (i.e., d-block elements). This reaffirms the characteristic that binary crystals with a larger difference in electronegativity between the two species typically have a larger bandgap. Interestingly, eight of the top fifteen binary compounds with the largest bandgaps were found to contain fluorine (the most electronegative element) as one of the elements. ii) The average bandgap generally reduces with the increased atomic number down a group, with materials consisting of at least one element belonging to periods 1-3 demonstrating a higher bandgap. The higher bandgap may be attributed to the smaller atomic radius of these elements that contributes to stronger bonds in the binary compound. Consequently, a larger amount of energy is required to move electrons to the conduction band, thus increasing the bandgap in these materials. It is remarkable that the D-ALIGNN model developed in this work learned these insights implicitly in automated protocols.

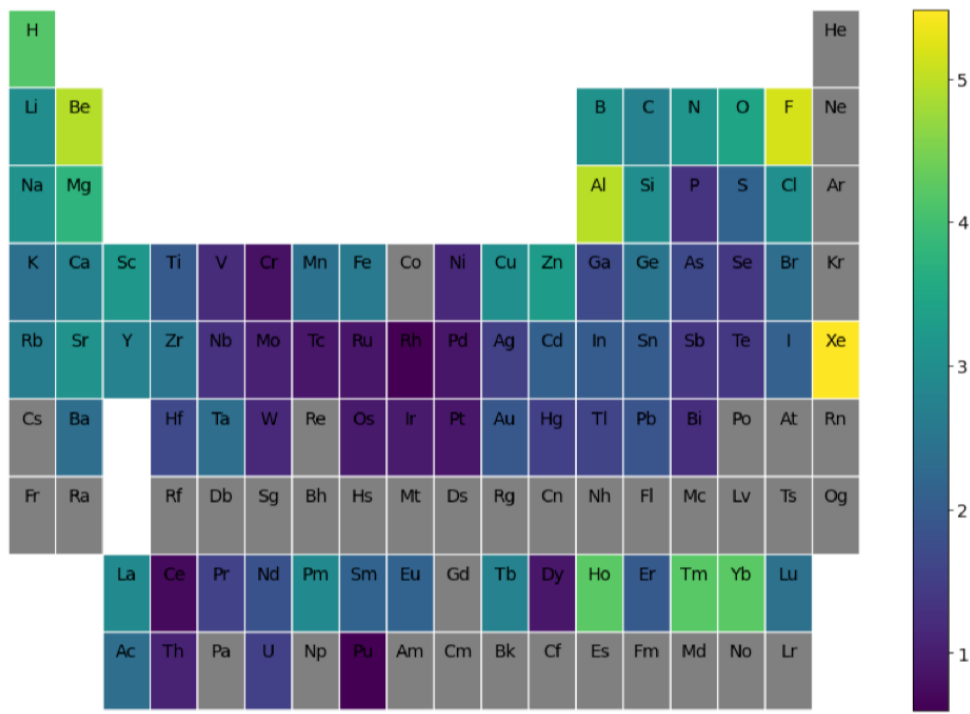

Figure 6. A heatmap of the periodic table, with elements shaded by the average bandgap of binary compounds containing them as computed from the DOS spectra predicted by D-ALIGNN.

**Conclusions**

In this study, we have proposed and evaluated the utility of ALIGNN models in predicting the DOS spectra of crystal structures. Specifically, the benefit of utilizing a graph-based featurization scheme that captures directional information is shown by demonstrating that the model can accurately predict the salient features of the DOS, which is a physical property that is inherently dependent on this information. We have also evaluated two different representational approaches for the DOS – a primitive discretization, and a compressed representation that is generated by using a separately trained autoencoder. Although the D-ALIGNN model performed better, both models exhibited sufficient accuracy to be used for high-throughput DOS spectrum predictions for new crystals. Most importantly, both modeling frameworks are scalable to include complex crystal systems that have varied structural and chemical diversity.

**Acknowledgements**

P.R.K. and S.R.K. gratefully acknowledge support from ONR N00014-18-1-2879. The Hive cluster at Georgia Institute of Technology (supported by NSF 1828187) was used for this work.

The authors declare that no known competing financial interests have influenced the work reported in this paper.

**Conflict of Interest**

The authors declare that there is no conflict of interest.